\documentclass[12pt]{article}
\usepackage[dvipdfmx]{graphicx} 
\usepackage{amsmath} 
\usepackage{amssymb} 
\usepackage{amsfonts} 
\usepackage{mathrsfs} 
\usepackage{leftidx} 
\usepackage{delarray} 
\usepackage{braket} 

\begin{document}

\begin{center}
 {\huge Generalized thermalization time of \\  \vspace*{2mm}  dark matter captured by neutron stars}
\end{center}
  \vspace*{2mm}
\begin{center}
{\large Yusuke Sanematsu and Motoi Tachibana}\\  \vspace*{2mm} Department of Physics, Saga University, Saga 840-8502, Japan
\end{center}


 \vspace*{3mm}
\begin{center}
{\large Abstract}
\end{center}
We discuss the issue on dark matter capture by neutron stars, in particular the process of dark matter thermalization, by which the scattering cross section and the mass of dark matter can be constrained. At first, we evaluate the thermalization time of self-interacting dark matter and find the effect of the self-interaction is small compared with that of the interaction with nucleons. Then we generalize the thermalization time by introducing a set of new parameters. We show how the cross section is affected by those new parameters. It turns out that
the cross section gets very sensitive to and strongly constrained by one of the new parameters. 
\noindent



\newpage

\section{Introduction}
There are enormous evidences for the existence of dark matter (DM) in the Universe [1]. For example,  according to the research of galaxy rotation curve, DM exists more than visible matters [2]. If DM is a weakly interacting massive particle (WIMP), it may interact with nucleons. Along this line, experiments for the direct detection of DM have been currently carried out on the Earth [3].  

On the other hand, the stellar constraints on the DM properties are remarkable as well. Among those, we consider
the issue on DM captured by neutron stars (NS) [4]. Since NS has strong gravity, once DM is captured inside of NS, it is hard to escape. In such a way, a huge amount of DM is accreted in NS and eventually gets self-gravitated. This is the onset of the gravitational instability, by which the host NS can be destructed. 

There are three major processes of physics described by the above [5]. First, DM is attracted by strong gravity of NS
and captured by the interactions with particles inside of NS. The DM capture rate is expressed as
\begin{align}
\frac{dN_\chi}{dt} = C_{\chi n} +C_{\chi \chi} N_\chi - C_a N_\chi^2, \label{eq:dndt}
\end{align}
where $N_\chi$ is the total number of DM in NS. Each term on the right hand side, with the coefficients 
$C_{\chi n}, C_{\chi \chi}$ and $C_a$, stems from interaction with nucleons, DM self-interaction and DM annihilation, respectively. 
Secondly, DM loses its kinetic energy through the interaction with nucleons. 
The DM energy loss rate is 
characterized by
\begin{align}
\frac{dE_\chi}{dt}= -\xi \sigma_{\chi n} v_\chi n_b \delta E_\chi \label{eq:dedt}
\end{align}
with the DM-nucleon cross section $\sigma_{\chi n}$, the DM velocity $v_\chi$, the nucleon number density $n_b$, and the DM energy loss per one scattering $\delta E_\chi$. $\xi$ denotes the effect of Pauli blocking. Using the Fermi momentum $p_F \simeq 426$ MeV, $\xi =\delta p/p_F =\sqrt{2}m_r v_\chi/p_F$. Here $m_r$ is the reduced mass composed of $m_\chi$ and $m_n$, mass of DM and nucleon, respectively. If DM loses sufficient energy, it gets thermalized within some finite region called the thermal radius, $r_{th}$, which is determined by the balance between the gravitational energy and the thermal kinetic energy. Then, more and more DMs are accumulated into the thermal radius and if the amount of DM exceeds the critical value, it has self-gravity. The self-gravity condition is given by
\begin{align}
\frac{3N_\chi m_\chi}{4\pi r_{th}^3} > \rho_b,
\end{align}
where $\rho_b$ is the nucleon mass density. It turns out that $r_{th}$ is related to the NS temperature, 
$T$, with the Newton's constant $G$ and the Boltzmann's constant $k$ as follows:
\begin{align}
r_{th} = \frac{3}{2} \sqrt{\frac{kT}{\pi G \rho_b m_\chi}}.
\end{align}

Here let us focus on the second stage, i.e., the DM thermalization process.  
The thermalization time can be estimated from eq.(2). To this end, let us assume that DM moves along the circular
orbit [6]. Then the DM kinetic energy $E_\chi$ is given by
\begin{align}
E_\chi = \frac{2 \pi}{3}G \rho_b m_\chi r_\chi^2 \label{eq:E.chi}
\end{align}
with the orbital radius $r_\chi$. Plugging this equation into (2), we have
\begin{align}
\frac{dr_\chi}{dt}= -\frac{4\sqrt{2}\pi G}{3p_F m_\chi} \sigma_{\chi n} r_\chi^3 n_b^2 m_r^3, \label{eq:drdt1}
\end{align}
which is easily solved as 
\begin{align}
r_\chi (t) = R\left( 1+\frac{8\sqrt{2}\pi G R^2}{3p_F m_\chi} \sigma_{\chi n} n_b^2 m_r^3 (t-t_1) \right)^{-1/2}.
\end{align}
Here $t_1$ is the time for the orbit to be contained within the NS radius $R$. 
By assuming $t_1$ is negligibly small compared with the thermalization time $t_{th}$, which is determined
by $r_{\chi}(t_{th})=r_{th}$, we obtain
\begin{align}
t_{th} \simeq \frac{(3\pi^2)^{1/3}}{6\sqrt{2}}  \frac{1}{kT}\left( \frac{m_\chi}{m_n} \right)^2 \left( \frac{m_n}{m_r} \right)^3 \left( \frac{1}{\sigma_{\chi n} n_b^{2/3}} \right). \label{eq:tth}
\end{align}

In this paper, the main purpose of our study is to generalize the above analysis
to a wider class of the DM models.
In section 2, we consider the effect of DM self-interaction on the thermalization. In section 3, we extend the previous studies and generalize eq.(\ref{eq:tth}) by introducing new parameters and see how those parameters affect the DM-nucleon scattering cross section. Section 4 is devoted to summary and conclusions.

\section{Effect of DM self-interaction on thermalization time}
In the previous section, DM self-interaction is not considered in the thermalization process. 
In this section, let us consider the case that DM loses its kinetic energy not only through the interaction 
with nucleons but also through the DM self-interaction. Then eq.(\ref{eq:dedt}) is modified as follows:
\begin{align}
\frac{dE_\chi}{dt} = \left( \frac{dE_\chi}{dt} \right)_{DM-n} + \left( \frac{dE_\chi}{dt} 
\right)_{DM-DM} \label{eq:E} , 
\end{align}
where the first term on the right hand side is due to the DM-nucleon scattering while the second one due to
the DM self-interaction.
They are expressed as 
\begin{align}
&\left( \frac{dE_\chi}{dt} \right)_{DM-n} = -\xi n_b \sigma_{\chi n} v_\chi \delta E_{DM-n} \label{eq:DMn},\\
&\left( \frac{dE_\chi}{dt} \right)_{DM-DM} = -n_\chi \sigma_{\chi \chi} v_\chi \delta E_{DM-DM} \label{eq:DMDM}
\end{align} 
with $\delta E_{DM-DM}=4\epsilon E_\chi $ \footnote[1]{For the derivation, see Appendix A.}. $\epsilon$ is defined through the restitution coefficient. $n_\chi$ and $\sigma_{\chi \chi}$ are the DM number density and the DM self-interaction cross section, respectively. Eqs.($\ref{eq:E}$), ($\ref{eq:DMn}$) and ($\ref{eq:DMDM}$) 
together with (\ref{eq:E.chi}) are combined to yield
\begin{align}
\frac{dr_\chi}{dt}=-Ar_\chi^3-Br_\chi^2, \label{eq:drdt2}
\end{align}
where
\begin{align}
A=\frac{4\sqrt{2}\pi G}{3p_F m_\chi} \sigma_{\chi n}  n_b^2 m_r^3, \qquad
B=-4\epsilon \sqrt{\frac{\pi}{3} G \rho_b} n_\chi \sigma_{\chi \chi}.
\end{align}
If $\epsilon$ is negligibly small, eq.(\ref{eq:drdt2}) reduces to eq.(\ref{eq:drdt1}). 
 Although $n_\chi$ actually depends on time, we regard $n_\chi$ as constant in time for simplicity. 
Then one can solve eq.(12) for $t_{th}$:
\begin{align}
t_{th} = -\frac{A}{B^2}\log{\left( 1+\frac{B}{Ar_{th}} \right) } + \frac{1}{Br_{th}}. \label{eq:tthab}
\end{align}
As was mentioned before, $t_{th}$ has to satisfy the condition $t_{th} < t_{NS}$.

Here we are interested in the effect of DM self-interaction. So let us see how $\sigma_{\chi n}$ changes as a function of $m_\chi$ and $\sigma_{\chi \chi}$ in eq.(\ref{eq:tthab}). To this end, we solve the stationary
conditions, i.e., $\partial \sigma_{\chi n}/ \partial m_\chi =0$, $\partial \sigma_{\chi n}/ \partial  \sigma_{\chi \chi} =0$, which lead us to
\begin{align}
m_\chi = \frac{1}{2} m_n.  \label{eq:mx}
\end{align}
For the derivation, see Appendix B.

\begin{figure}[htbp]
  \begin{center}
    \includegraphics[clip,height=10.0cm, width=10.0cm]{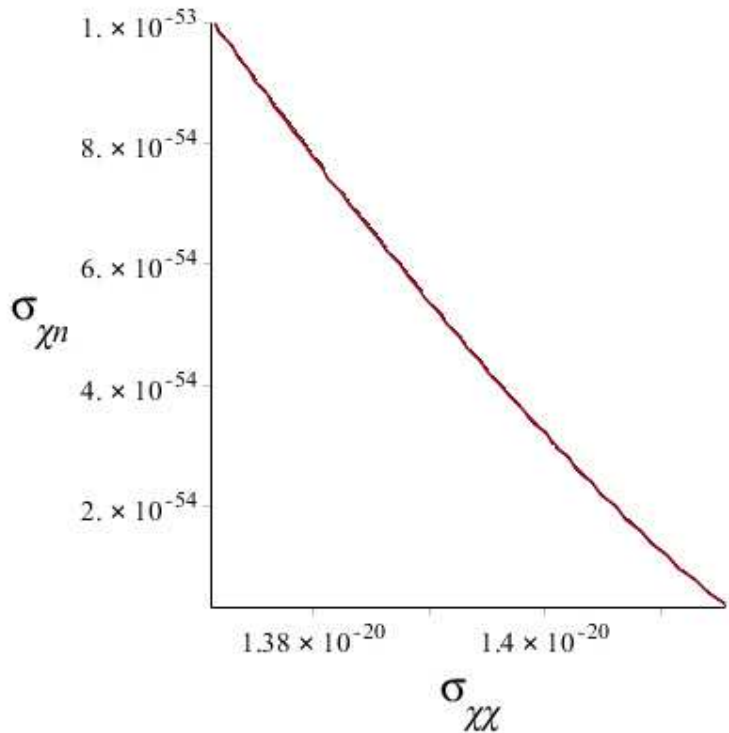}
    \caption{Relation between $\sigma_{\chi n}$ and $\sigma_{\chi \chi}$ for $m_\chi = \frac{1}{2} m_n$}
    \label{fig:sigma}
  \end{center}
\end{figure}

By plugging eq.(\ref{eq:mx}) into eq.(\ref{eq:tthab}) and putting $t_{th}=10^{10}$ years (the oldest age of the observed NS), $n_{\chi}=10^{3} \ \rm{cm^{-3}}$ (DM number density around the sun), $n_b=4.0\times 10^{38} \ \rm{cm^{-3}}$ (a typical value of baryon number density in NS) and $\epsilon =0.5$, we obtain $\sigma_{\chi n}$ as a function of $\sigma_{\chi \chi}$. 
The result is shown in Figure \ref{fig:sigma}.



 As you see from Figure 1, $\sigma_{\chi n}$ hardly changes by $\sigma_{\chi \chi}$. 
So we find that the effect of the self-interaction on the DM thermalization in NS
is negligibly small. Thus, we do not take into account the effect below.
 

\section{Generalized thermalization time }

In the previous section, we discussed the effect of the DM self-interaction and found that it can be neglected. 
So below we concentrate on the DM scattering with nucleons and try to extend the analysis previously done.

\subsection{Introduction of new parameters}
For the purpose, we introduce a set of new parameters $(\alpha, \beta, \gamma)$ 
to generalize the thermalization time formula eq.(\ref{eq:tth}) as follows:
\begin{align}
t_{th} &=\frac{(3\pi^2)^{1/3}}{6\sqrt{2}} \frac{1}{kT} \left( \frac{1}{\sigma_{\chi n} {n_b}^{2/3}} \right)^\alpha \left( \frac{m_\chi}{m_n} \right)^\beta \left( \frac{m_n}{m_r} \right)^\gamma  \nonumber \\
&=C\left( \frac{1}{\sigma_{\chi n} {n_b}^{2/3}} \right)^\alpha X^\beta \left( 1+\frac{1}{X} \right)^\gamma\label{eq:tth2}
\end{align}
with
\begin{align}
X=\frac{m_\chi}{m_n}, \qquad \quad
C=\frac{(3\pi^2)^{1/3}}{6\sqrt{2} kT}. 
\end{align}
$(\alpha, \beta, \gamma)=(1, 2, 3)$ corresponds to the original one, eq.(\ref{eq:tth}).
Taking the log of both sides of eq.($\ref{eq:tth2}$) reads
\begin{align}
\log_{10}{t_{th}} = -\alpha \log_{10}{\sigma_{\chi n}} + \beta \log_{10}{X} + \gamma \log_{10}{(1+\frac{1}{X})} +C_1 \label{eq:log},
\end{align}
where $C_1=\log_{10}(C/{n_b^{2 \alpha /3}})$.

For $m_\chi \gg m_n \ (X \gg 1)$, eq.($\ref{eq:log}$) is approximated by
\begin{align}
\log_{10}{\sigma_{\chi n}} \simeq \frac{\beta}{\alpha} \log_{10}{X} + \frac{1}{\alpha} C_2,
\end{align}
where $C_2=\log_{10}[C/({n_b^{2 \alpha /3}t_{th}})]$.

For $m_\chi \ll m_n \ (X \ll 1)$, eq.($\ref{eq:log}$) is approximated by
\begin{align}
\log_{10}{\sigma_{\chi n}} \simeq \frac{\beta - \gamma}{\alpha} \log_{10}{X} + \frac{1}{\alpha} C_2. 
\end{align}
In order to obtain the extreme value of $\sigma_{\chi n}$, eq.(\ref{eq:tth2}) is differentiated by $X$. 
Then the extreme value $X_{ext}$ is obtained as
\begin{align}
X_{ext} = \frac{\gamma - \beta}{\beta}. \label{eq:xp}
\end{align}
Since $X_{ext}$ is the fraction of mass, it must be positive. Therefore the following inequalities should be fulfilled (Figure 2).
\begin{align}
&\beta < \gamma \qquad  {\rm for} \qquad \beta , \gamma > 0, \\
&\beta > \gamma \qquad  {\rm for} \qquad \beta , \gamma < 0. 
\end{align}
 For $\beta < \gamma$, the shape of $\sigma_{\chi n}$ as a function of $m_{\chi}$ is  concave up while for $\beta > \gamma$, it is concave down. Moreover, since $X_{ext}$ should be in the middle of range between $X\gg 1$ and 
 $X\ll 1$, it has to obey the following relation:
\begin{align}
X_{ext} = \frac{\gamma - \beta}{\beta} \simeq {\cal O}(1)  \quad
\longrightarrow \quad \frac{\gamma}{\beta} \simeq {\cal O}(1). \label{eq:beta}
\end{align} 
Thus the values of $\beta$ and $\gamma$ is nearby, i.e., $\gamma \simeq \beta$.\\
\begin{figure}[htbp]
  \begin{center}
    \includegraphics[clip,height=8.5cm, width=9.5cm]{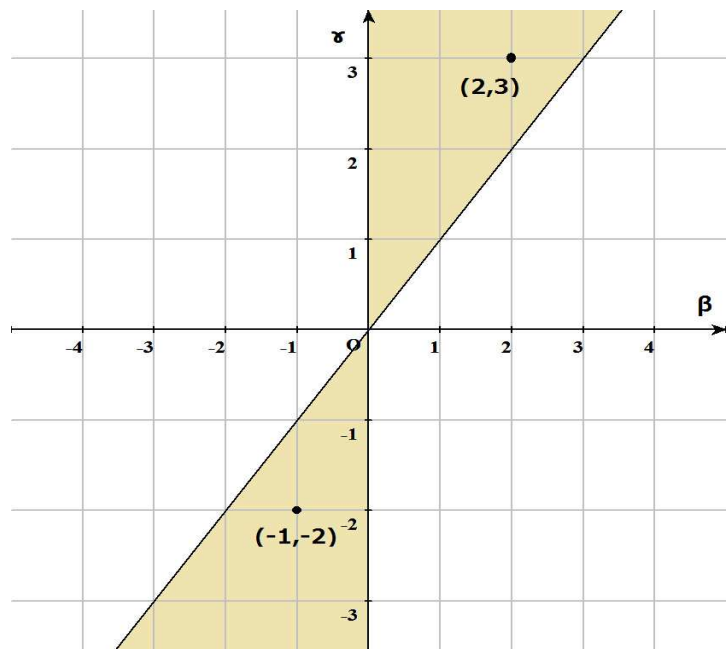}
    \caption{Allowed region on $\beta - \gamma$ plane}
    \label{fig:gamma}
  \end{center}
\end{figure}

Note here that in the generalized thermalization time formula (16), 
the results of preceding studies [5] and [7] correspond to
two different parameter sets $(\beta , \gamma) = (2,3)$ and $(\beta , \gamma) = (-1,-2)$, respectively. 
Those are plotted in Figure 2.

Using (\ref{eq:tth2}) and (\ref{eq:xp}), 
the extreme value of cross section, $\sigma_{\chi n}^{ext}$ is given by
\begin{align}
\sigma_{\chi n}^{ext} = 1.8 \times 10^{-26} \left[ 8.4\times 10^{-35} \frac{\gamma^\gamma}{\beta^\beta} \left( \frac{1}{\gamma -\beta} \right)^{\gamma - \beta} \right]^{1/\alpha}.
\end{align}
Here we put $t_{th} = 10^{10}$ years, $T = 10^5$ K and $n_b=4.0\times 10^{38} \ \rm{cm^{-3}}$.
From the above equation, we expect that the change of $\alpha $ largely affects the value of $\sigma_{\chi n}^{ext}$, compared to $\beta$ and $\gamma$. In the following, let us see it more concretely.


\subsection{$\alpha$ dependence of $\sigma_{\chi n}$}
Since we are interested in the $\alpha$ dependence of $\sigma_{\chi n}$, 
let us keep $\beta$ and $\gamma$ to the original value, namely $(\beta, \gamma)=(2, 3)$. 
Then eq.(\ref{eq:tth2}) becomes
\begin{align}
t_{th} =C \left( \frac{1}{\sigma_{\chi n} n_b^{2/3}} \right)^\alpha X^2 \left( 1+\frac{1}{X} \right)^3. \label{eq:alpha}
\end{align}
Similarly in the previous subsection, we evaluate the extreme value of $\sigma_{\chi n}$. 
By solving $\partial \sigma_{\chi n} / \partial X =0$, we have
\begin{align}
X_{ext}=\frac{1}{2} \quad
\longrightarrow \quad m_\chi = \frac{1}{2} m_n. \label{eq:mchi}
\end{align}
The result is independent with $\alpha$. It means that whenever eq.(\ref{eq:mchi}) holds, $\sigma_{\chi n}$
gets the extreme value.

Let us now see how the value of $\sigma_{\chi n}$ changes for some different values of $\alpha$. 
For $\alpha = 1$, the extreme value of $\sigma_{\chi n}$ is approximately given by
\begin{align}
\sigma_{\chi n}^{\alpha =1} \approx 1.1 \times 10^{-59} \  [\rm{cm^2}].
\end{align}
Similarly, for $\alpha =\frac{1}{2}$, $\alpha =2$ and $\alpha =3$, the following results are obtained:
\begin{align}
\sigma_{\chi n}^{\alpha=1/2} &\approx 6.5 \times 10^{-93} \ [\rm{cm^2}],
\end{align}
\begin{align}
\sigma_{\chi n}^{\alpha=2} &\approx 4.5 \times 10^{-44} \ [\rm{cm^2}],
\end{align}
\begin{align}
\sigma_{\chi n}^{\alpha=3} &\approx 1.6 \times 10^{-37} \ [\rm{cm^2}].
\end{align}
See Figure 3. Therefore, for larger $\alpha$, the extreme value of $\sigma_{\chi n}$ becomes larger. 
From this, one can conclude that the DM-nucleon scattering cross section is very sensitive to
and strongly constrained by $\alpha$.
Comparing with results from the direct detection experiments, $\alpha=3$
seems to be already excluded. 


\begin{figure}[htbp]
  \begin{center}
\begin{tabular}{c}    

      \begin{minipage}{0.50\hsize}
        \begin{center}
          \includegraphics[clip, width=7.0cm]{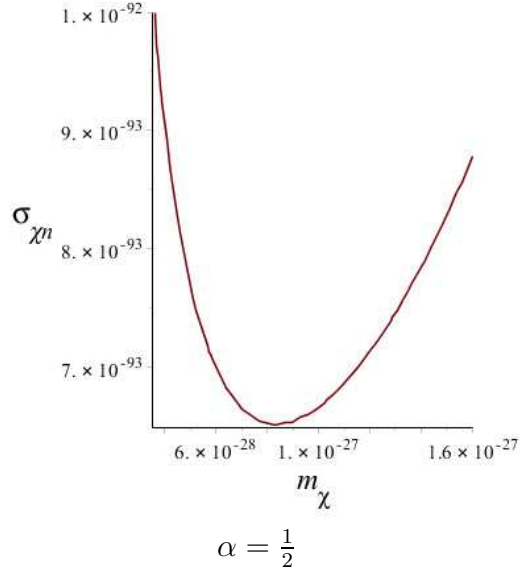}
          \hspace{2.0cm} $\alpha=\frac{1}{2}$
        \end{center}
      \end{minipage}

      \begin{minipage}{0.50\hsize}
        \begin{center}
          \includegraphics[clip, width=7.0cm]{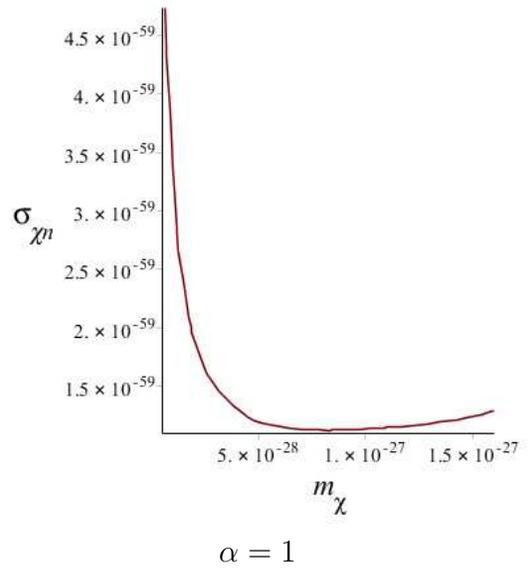}
          \hspace{2.0cm} $\alpha=1$
        \end{center}
      \end{minipage}
\\
\\
\\
\\

\end{tabular}  

\begin{tabular}{c}
      \begin{minipage}{0.50\hsize}
        \begin{center}
          \includegraphics[clip, width=7.0cm]{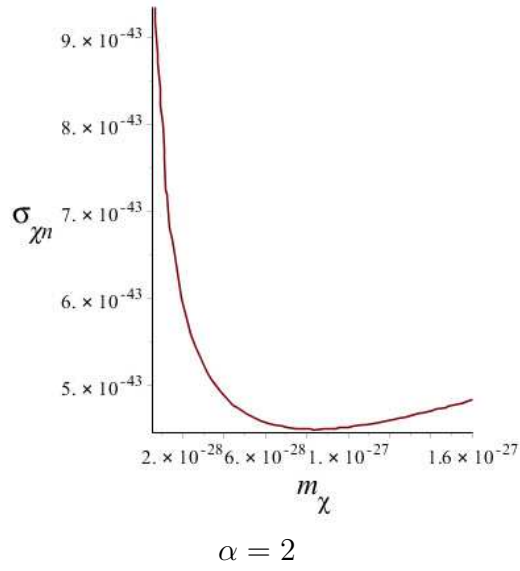}
          \hspace{2.0cm}$\alpha = 2$
        \end{center}
      \end{minipage}

      \begin{minipage}{0.50\hsize}
        \begin{center}
          \includegraphics[clip, width=7.0cm]{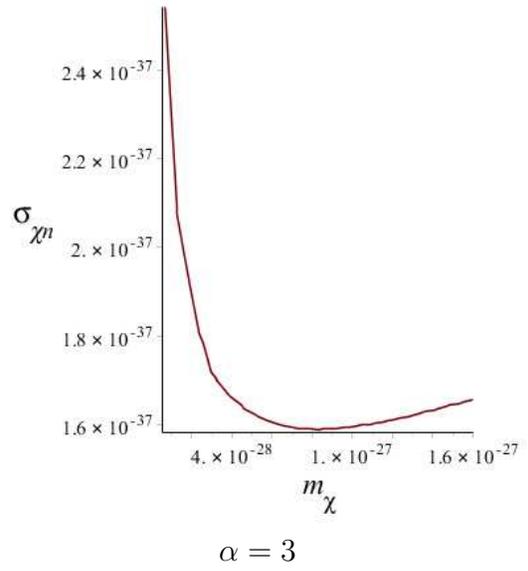}
          \hspace{2.0cm}$\alpha =3$
        \end{center}
      \end{minipage}
\end{tabular}

    \caption{$\alpha$ dependence of the $\sigma_{\chi n}$-$m_{\chi}$ plot}
    \label{fig:a}
  \end{center}
\end{figure}

\newpage
\section{Conclusions}

In this paper, we discussed the issue on dark matter thermalization in neutron stars. We studied
two aspects, which were not taken into account previously. One is to consider
the effect of DM self-interaction to compute the thermalization time. We found that the effect is very small. 
Then we generalized the thermalization time formula by introducing new parameters. We found how the DM-nucleon
scattering cross section is affected by those parameters and in particular, for one of the parameters ($\alpha$),
the cross section is very sensitive and strongly constrained.

There are several comments. Here, we treated the DM number density $n_\chi$ as constant in time.
However, $n_\chi$ is actually the function of time and gets increased as time goes by.
Also we only focused on the extreme value of $\sigma_{\chi n}$ in our analysis. This would not be enough
to conclude that the effect of the DM self-interaction is negligible. We found that the parameter $\alpha$
has a sizable effect when it becomes larger than 1, but so far we do not know any specific model.
Moreover, we have only considered the case that DM interacts with normal nucleons. But in NS,
we could have some exotic phases such as quark matter as well as
superfluid and/or superconducting nucleons and they might play roles [8].
These are our future works.



 Besides, from the requirement of black hole formation as well as the gravitational instability mentioned in Introduction, the DM properties such as the cross section and the mass could be more constrained. 
 This is also an interesting future prospect of our research.

\section*{Acknowledgments}
The authors thank the Yukawa Institute for Theoretical Physics at Kyoto University. Discussions during the YITP workshop YITP-W-15-09 on "Thermal Quantum Field Theory and its application" were useful to complete this work.

\section*{Appendix A}
Let us consider two dark matters (DM1 and DM2) with the mass $m_\chi$.
DM1 and DM2 collide with initial velocities $v_1$ and $v_2$. After the collision, their velocities
are $v_1'$ and $v_2'$, respectively.
 Then one can define the restitution coefficient $e\  (0\leq e \leq 1)$ as follows:
\begin{align}
e=-\frac{v_1' -v_2'}{v_1-v_2}.
\end{align}  
The energy and the momentum conservation laws tell us
\begin{align}
m_\chi v_1+m_\chi v_2 &= m_\chi v_1' +m_\chi v_2', \\
\frac{1}{2}m_\chi v_1^2 +\frac{1}{2}m_\chi v_2^2&=\frac{1}{2}m_\chi v_1'^2 +\frac{1}{2}m_\chi v_2'^2 +\delta E_\chi. \label{eq:energy}
\end{align}
The solutions of these equations are
\begin{align}
v_1' = \frac{1}{2} \Big( v_1 \left( 1-e \right) + v_2 \left( 1+e \right) \Big) \label{eq:v1} \\
v_2' = \frac{1}{2} \Big( v_1 \left( 1+e \right) + v_2 \left( 1-e \right) \Big). \label{eq:v2}
\end{align}
By assuming $v_2=-v_1$, we get $v_1' =-ev_1$, $v_2'=v_1$ from eq.(\ref{eq:v1}) and eq.(\ref{eq:v2}). 
As the result,
\begin{align}
\delta E_\chi =m_\chi v_1^2 (1-e^2) =2E_\chi(1-e^2).
\end{align}
Here $E_\chi$ is the DM kinetic energy. If we adopt $e=1-\epsilon$ and assume that $\epsilon \ll 1$, then
\begin{align}
\delta E_\chi = 2E_\chi (2\epsilon -\epsilon^2) \simeq 4\epsilon E_\chi.
\end{align}

\section*{Appendix B}
The partial differentiation of eq.(\ref{eq:tthab}) by $m_\chi$ leads that
\begin{align}
0&= \frac{\partial A}{\partial m_\chi} \left( -\frac{1}{B^2} \log{\left( 1+\frac{B}{A r_{th}}\right)} +\frac{1}{B(Ar_{th}+B)} \right) \nonumber \\
 &+ \frac{\partial r_{th}}{\partial m_\chi} \left( \frac{A}{Br_{th}(Ar_{th}+B)} - \frac{1}{Br_{th}^2} \right) \label{eq:A}.
\end{align}
Similarly by the partial differentiation of eq.(\ref{eq:tthab}) with respect to $\sigma_{\chi \chi}$, 
\begin{align}
0=\frac{\partial B}{\partial \sigma_{\chi \chi}} \left( \frac{2A}{B^3}\log{\left( 1+\frac{B}{Ar_{th}}\right)} +\frac{Ar_{th}}{B^2 r_{th}(Ar_{th} +B) } -\frac{1}{B^2 r_{th}} \right)\label{eq:B}.
\end{align}
Clearly from the definitions of $A$, $B$ and $r_{th}$, $A$ is the function of $\sigma_{\chi n}$ and $m_\chi$, $B$ the function of $\sigma_{\chi \chi}$, and $r_{th}$ the function of $m_\chi$. 
Since $\frac{\partial B}{\partial \sigma_{\chi \chi}} \neq 0$ in eq.(\ref{eq:B}), 
we easily find
\begin{align}
\frac{2A}{B^3}\log{\left( 1+\frac{B}{Ar_{th}}\right)} &+\frac{Ar_{th}}{B^2 r_{th}(Ar_{th} +B) } -\frac{1}{B^2 r_{th}}=0 \nonumber \\
\longrightarrow \log{\left( 1+\frac{B}{Ar_{th}} \right)} &= \frac{B}{2(Ar_{th}+B)} + \frac{B}{2Ar_{th}}.\label{eq:B2}
\end{align}
By plugging (\ref{eq:B2}) into (\ref{eq:A}),
\begin{align}
& 0=r_{th}\frac{\partial A}{\partial m_\chi} +2A\frac{\partial r_{th}}{\partial m_\chi} \nonumber \\
 \longrightarrow \quad & 0= m_{\chi}\frac{\partial \sigma_{\chi n}}{\partial m_\chi}+
\sigma_{\chi n}\frac{2m_n - m_\chi}{m_n +m_\chi} -\sigma_{\chi n} \label{eq:0}. 
\end{align}
If we require $\frac{\partial \sigma_{\chi n}}{\partial m_\chi}=0$ in eq.(\ref{eq:0}), we obtain
\begin{align}
m_\chi =\frac{1}{2} m_n.
\end{align}


\begin{thebibliography}{99}
\bibitem{24.DM - Particle Data Group (2013)} Particle Data Group 24. Dark Matter (M.Drees and G. Gerbier). 
\bibitem{rotation curve} F. Zwicky, Helvetica Physica Acta 6 (1933) 110-127.\\
F. Zwicky, Astrophys. J. 86 (1937) 217.
\bibitem{direct detection} For instance, see http://xenon.astro.columbia.edu/. (the Xenon project)\\
http://cdms.berkeley.edu/. (the CDMS project)
\bibitem{DM in NS} I. Goldman and S. Nussinov, Phys. Rev. D40 (1989) 3221-3230.\\
C. Kouvaris, Phys. Rev. D77 (2008) 023006.
\bibitem{Zurek2012}S. D. McDermott, H-B. Yu and K. Zurek, Phys. Rev. D85 (2012) 023519.
\bibitem{On the capture of DM by NS}T. G\"{u}ver et al., JCAP 1405 (2014) 013. 
\bibitem{Reddy2013} B. Bertoni, A. Nelson and S. Reddy, Phys. Rev. D88 (2013) 123505.
\bibitem{Tachibana2013,2015} M. Ruggieri and M. Tachibana, arXiv:1312.5802 [hep-ph].\\
M. Tachibana, to appear in Acta Astronomica Sinica 56 (2015).
\end{thebibliography}
\end{document}